\documentclass[useAMS,usenatbib]{mn2e}

\usepackage{graphicx,amssymb}
\usepackage[fleqn]{amsmath}

\newcommand{\bb}{\mbox{\boldmath $b$}}
\newcommand{\vr}{\mbox{\boldmath $r$}}
\newcommand{\vO}{\mbox{\bf 0}}

\newcommand{\er}{\mbox{\boldmath $\hat{e}$}_r}
\newcommand{\et}{\mbox{\boldmath $\hat{e}$}_t}
\newcommand{\be}{\begin{equation}}
\newcommand{\ee}{\end{equation}}
\newcommand{\ba}{\begin{eqnarray}}
\newcommand{\ea}{\end{eqnarray}}

\newcommand{\com}[1]{{#1}}

\title[Cluster Bulleticity]{Cluster Bulleticity}
\author[Massey, Kitching \& Nagai]{Richard Massey$^{1}$, Thomas Kitching$^{1}$ \& Daisuke Nagai$^{2,3}$\\
$^{1}$
University of Edinburgh, Royal Observatory, Blackford Hill, Edinburgh EH9 3HJ, UK\ \ {\tt rm},{\tt tdk@roe.ac.uk}\\
$^{2}$Department of Physics, Yale University, New Haven, CT 06520, USA\ \ {\tt daisuke.nagai@yale.edu}\\
$^{3}$Yale Center for Astronomy \& Astrophysics, New Haven, CT 06520, USA}
\begin{document}


\pagerange{\pageref{firstpage}--\pageref{lastpage}} \pubyear{2010}

\maketitle

\label{firstpage}

\begin{abstract}
The unique properties of dark matter are revealed during 
collisions between clusters of galaxies, like
the bullet cluster (1E 0657-56) and baby bullet (MACSJ0025-12). 
These systems provide evidence for an additional, invisible mass
in the separation between the distribution of their total mass, 
measured via gravitational lensing, and their ordinary `baryonic' 
matter, measured via its X-ray emission.
Unfortunately, the information available from these systems is
limited by their rarity. Constraints on the properties of 
dark matter, such as its interaction cross-section, are therefore 
restricted by uncertainties in the individual systems' impact velocity, 
impact parameter and orientation with respect to the line of sight.

Here we develop a complementary, statistical measurement in which every piece of
substructure falling into every massive cluster is treated as a bullet. We
define `bulleticity' as the mean separation between dark matter and ordinary
matter, and we measure \com{the} signal in hydrodynamical simulations. The phase
space of substructure orbits also exhibits symmetries that provide
\com{an equivalent control test.}

Any detection of bulleticity in real data would indicate a difference in the
interaction cross-sections of baryonic and dark matter that may rule out
hypotheses of non-particulate dark matter that are otherwise able to model
individual systems. \com{A subsequent measurement of bulleticity could constrain
the dark matter cross-section. Even with conservative estimates, the existing
HST archive should yield an independent constraint tighter than that from the
bullet cluster. This technique is then trivially extendable to, and benefits
enormously from larger, future surveys.}

%

\end{abstract}


\section{Introduction}

The standard $\Lambda$CDM cosmological model includes a component of cold dark
matter that amounts to 85\% of the matter content of the Universe. This dark
matter \com{affects the Universe primarily through gravity, and is necessary to
explain the distribution and growth of large scale structure over cosmic time.
However, dark matter does not interact (or only very weakly) in the electroweak
sector.}

The fundamentally different properties of dark matter and baryonic matter are
highlighted \com{most dramatically} in their temporary separation during
collisions between galaxy cluster pairs, such as 1E~0657-56
\citep{clowe04,clowe06,bradac06} and MACSJ0025-12 \citep{bradac08}.
These `bullet clusters' have provided astrophysical constraints on
the interaction cross-section \com{$\sigma$} of hypothesised dark matter particles, and may
ultimate prove the most useful laboratory in which to test for any velocity
dependence of the cross section. Unfortunately, the \com{utility of a small
number of} individual systems is limited by \com{observational} uncertainties in
their collision velocity, impact parameter and angle with respect to the plane
of the sky \citep{randall08}. Current constraints are 3 orders of magnitude
weaker than constraints from the shapes of haloes \citep{feng10} and, since
collisions between two massive progenitors are rare \citep{shan10}, the total
observable number of such systems may be inadequate to investigate a physically
interesting regime of dark matter properties.

%

In this paper, we present a statistical method that allows every piece of
substructure falling into every cluster to contribute to a global measure of
dark matter-baryonic separation that we refer to as `bulleticity'. In this
approach, every infalling mass is treated as a bullet, whose (interacting) gas
is expected to collide with the intra-cluster medium (ICM) and lag behind the
(non-interacting) dark matter \citep{powell09}. Although offsets between
baryonic gas (e.g.\ as seen in X-ray emission) and total mass (e.g.\ as seen by
gravitational lensing) may not be \com{{\it individually} significant, detecting
a mean bulleticity across many systems would provide robust} evidence for a
difference between the baryonic matter and dark matter interaction
cross-sections. \com{Measurements of the {\it amplitude} of bulleticity could
then constrain the level of the dark matter-dark matter and dark matter-baryonic
cross-sections.} Crucially, since \com{bulleticity} should be observable in the
ongoing assembly of every massive structure throughout the Universe, our
statistical technique can overcome the previous limitations of small number
statistics.


This paper is organised as follows. In Section~\S\ref{sec:toymodel}, we develop
a simple physical model to illustrate the concept of bulleticity and explore
some of its dependencies. In Section~\S\ref{sec:hydro}, we use full
hydrodynamical calculations to measure the expected bulleticity signal in
realistic galaxy clusters. \com{In Section~\S\ref{sec:practicality}, we discuss
the practicality of measurements from real astronomical data and use a Fisher
matrix analysis to predict constraints on the interaction cross-section of dark
matter using various data sets. We conclude} in Section~\S\ref{sec:discuss}.

\section{Dynamical model}\label{sec:toymodel}

\subsection{Motion of test particles falling into clusters}

We shall illustrate the dominant physical effects that separate components of
dark and baryonic mass as they fall into a massive cluster. 
Throughout, we shall refer to the combined infalling system as a `bullet'. 

Consider the rest frame of a cluster with fixed `singular isothermal sphere'
3D mass distribution $\rho(r)=\rho_0(r_0/r)^{2}$, which has a mass interior to 
radius $r$ of $M(<r)=4\pi\rho_0r_0^2r$.
An infalling, point-like component at position $\vr$, \com{which} interacts only
gravitationally, has an equation of motion
\begin{equation}
\label{dm}
\frac{{\mathrm d}^2\vr}{{\mathrm d} t^2}~~=~~-\frac{4\pi G\rho_0r_0^2}{r^2}\vr~,
\end{equation}
where $t$ is the proper time for the cluster and $r$ is the 3D radius. 

\com{An equivalent component that also interacts via the electroweak force}
experiences additional pressure support.
We approximate this \com{interaction} as a buoyancy force 
\com{equal to the weight of the displaced mass.}
For a sufficiently small bullet that
the density $\rho(r)$ of the cluster 
is constant across it, the 
equation of motion is
\begin{equation}
\frac{{\mathrm d}^2\vr}{{\mathrm d} t^2} ~~ \approx ~~
-\frac{4\pi G\rho_0r_0^2}{r^2}\left(1-(137\alpha)^2\frac{\rho_0r_0^2}{\rho_br^2}\right)\vr~,
\label{bm}
\end{equation}
where \com{$\alpha$ is the dimensionless coupling constant and 
$\rho_b$ is the mean (total) density of the bullet, assuming that the
ratio of baryons to dark matter is the same in the cluster and the bullet.

We shall use equation~(\ref{dm}) with $\alpha=0$ to model the dynamics of
standard cold dark matter and equation~(\ref{bm}) with $\alpha=1/137$ to model
baryonic matter. To study interacting dark matter, we can simply add a nonzero
$\alpha$ term, which represents the mean of the dark matter-dark matter and dark
matter-baryonic coupling constants, weighted by the ratio in which the two forms
of matter are found in the cluster. Such particles would follow orbits between
those of standard dark matter and baryons. The $(137\alpha)^2$ prefactor can be
interpreted as either the fractional volume of cluster mass displaced by a solid
bullet, or as the fractional cross-section seen by an interacting  particle,
i.e.\ for a geometrically thin bullet 
$(137\alpha)^2\approx\sigma/\pi r_b^2$.}

\begin{figure}
 \includegraphics[width = 230pt]{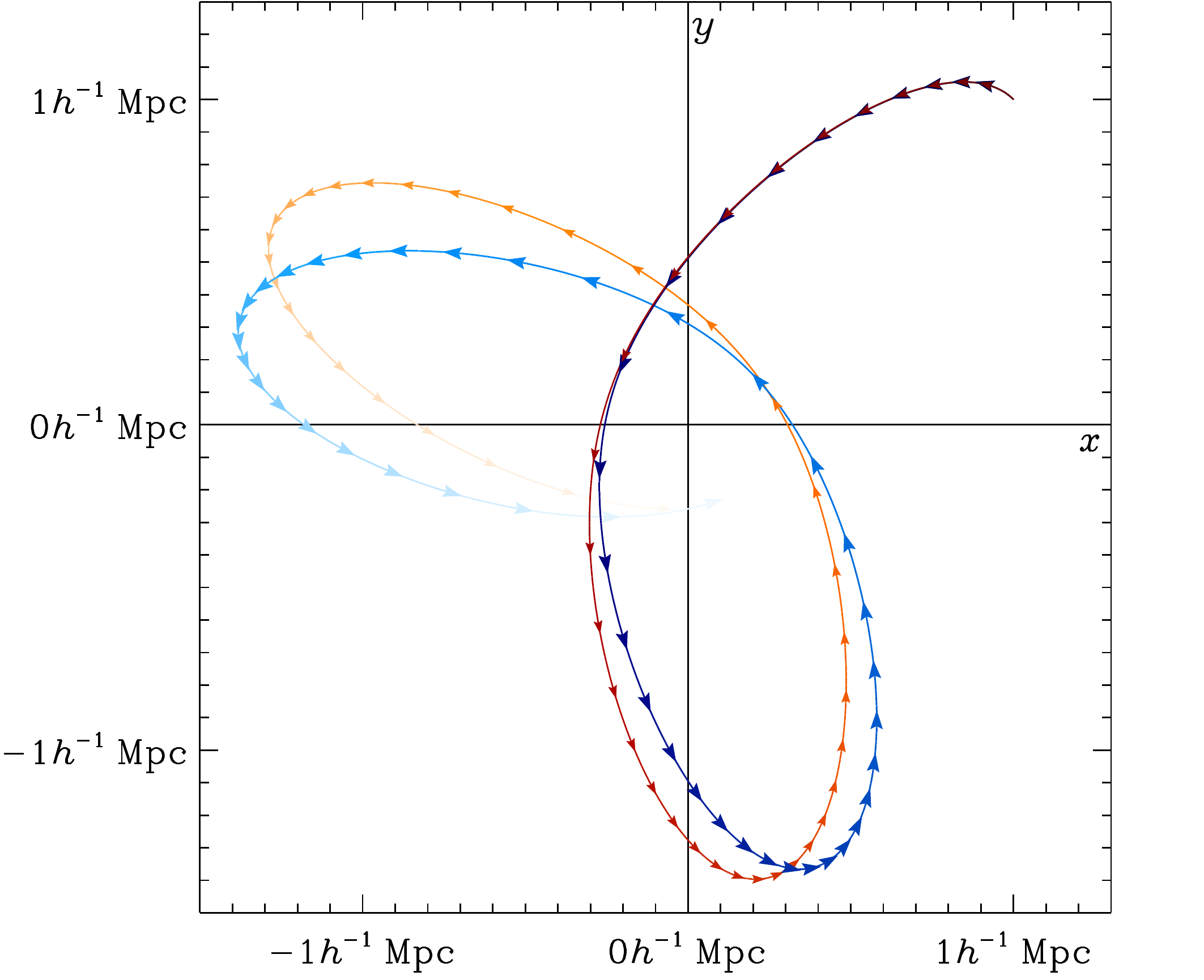}
 \includegraphics[width = 230pt]{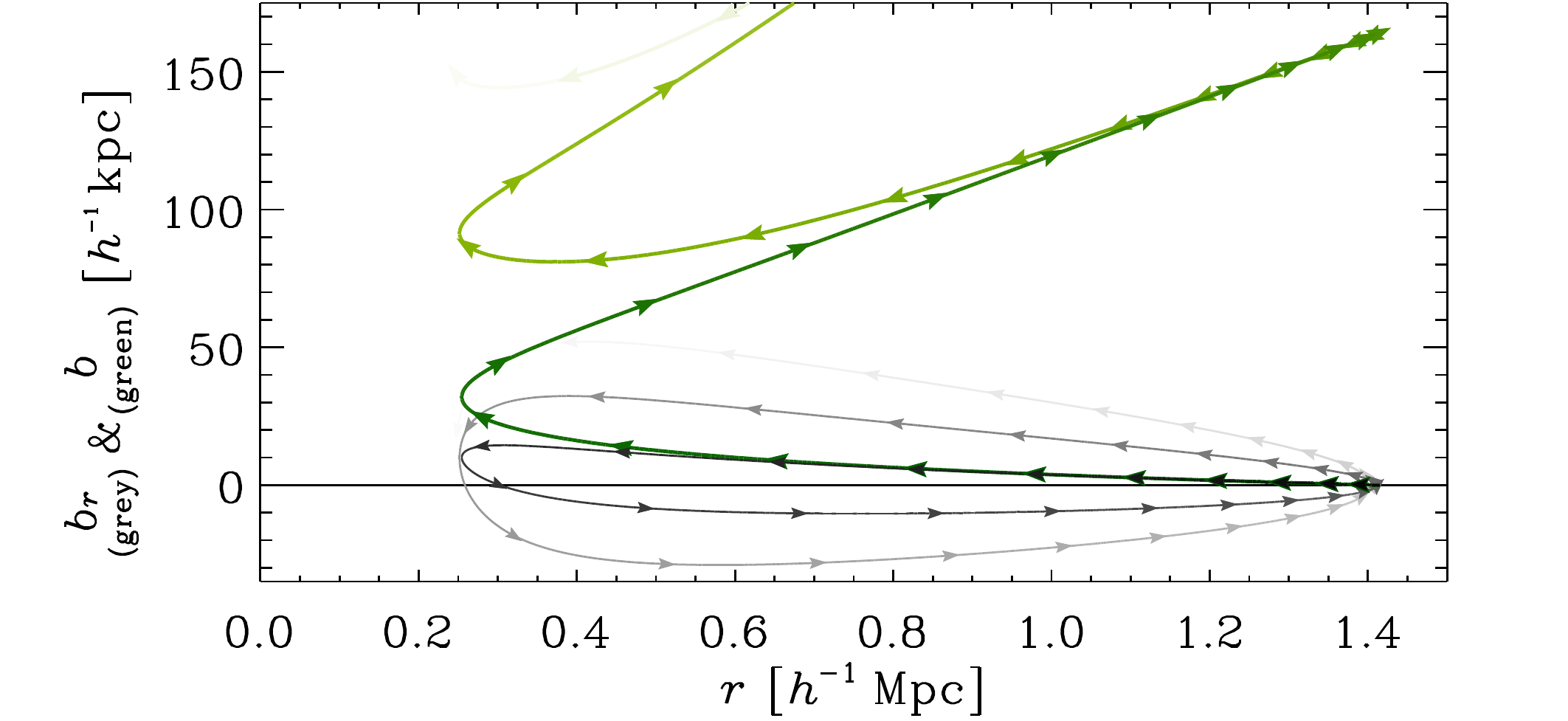}
 \caption{Orbits of test particles falling into a $10^{14}\,M_\odot$ 
 `singular isothermal sphere' cluster.
 In the top panel, the  blue curve (large arrows) shows the trajectory of 
 standard, non-interacting \com{$\alpha=0$} cold dark
 matter, which is governed in this simple model purely by the gravitational
 attraction from the central mass. The red curve (small arrows) shows the
 trajectory of \com{$\alpha=1/137$} baryonic matter, which experiences an 
 additional buoyancy force. 
 \com{Arrowheads are spaced uniformly in proper time, and the paths progressively fade.}
 The bottom panel shows the apparent bulleticity if the substructure is
 moving in the plane of the sky. \label{fig:orbit}}
\end{figure}

Note that both dark and baryonic matter  also experience dynamical friction and
tidal gravitational forces, but equally -- so that neither \com{of these
effects} should separate the two components. To keep our model simple, we
therefore ignore these effects, but note that the dissipation of the bullet due
to tidal and ram-pressure stripping, in conjunction with the finite crossing
time of a typical cluster, means that we would only ever expect to see bullets
on their first one or two passes through a cluster. Whilst the dynamical
friction of a lumpy intra-cluster medium would begin to circularise the bullet's
orbit during this time, we also neglect this effect.


\subsection{Definition of bulleticity}

We define bulleticity $\bb$ as the vector from the position of the dark matter,
projected onto the plane of the sky, to that of the baryons. \com{Its magnitude
is thus sensitive to the difference $\Delta\sigma$ in the total interaction
cross-sections of dark matter and baryons. Were all matter} to have the same
interaction cross-section, this would ensure $\bb\equiv\vO$. Any non-zero
detection will therefore indicate \com{the presence of nonbaryonic material.}

We define the location of the bullet to be the mean position $(r,\theta)$ of the
dark matter and the baryons projected onto the sky. \com{In general, these need not be the same. We
need to uniquely define the location of each bullet because its bulleticity can
depend upon the length of time it has been within a cluster, and the path it has
previously traversed.} Henceforth, we shall use the symbol $r$ to represent this 2D projected radius.


In the 1D case of a bullet that falls into a cluster potential along a radius,
equations~(\ref{dm}) and (\ref{bm}) both produce oscillatory motion (with the
caveat of a numerical finesse at the origin to remove the point of infinite
density). However, as the baryons experience extra buoyancy, they gradually lag
behind the dark matter: farther from the cluster core during
infall and closer to the core during egress. The absolute bulleticity
$b=|\bb|$ \com{steadily} increases.

\begin{figure}
 \includegraphics[width = 230pt]{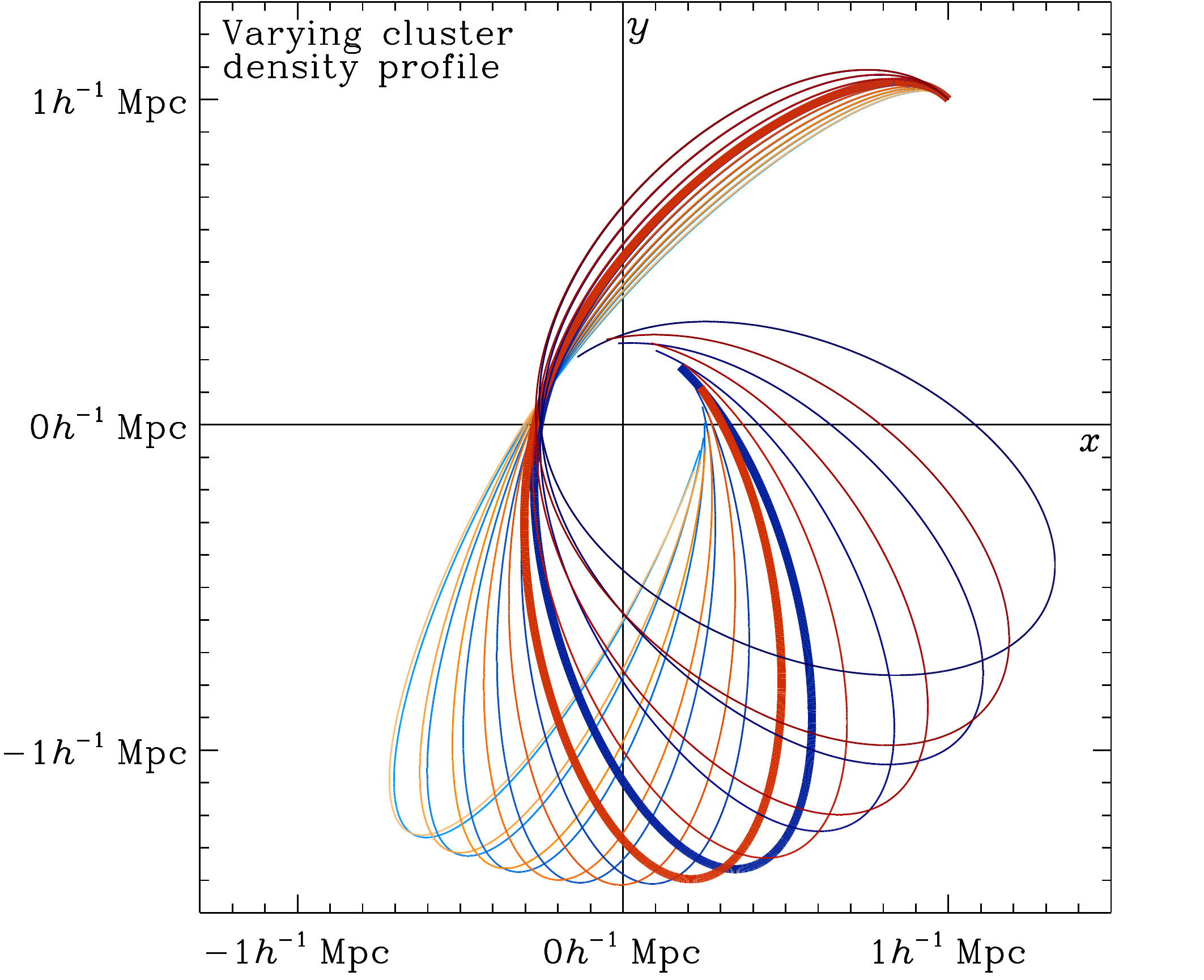}
 \includegraphics[width = 230pt]{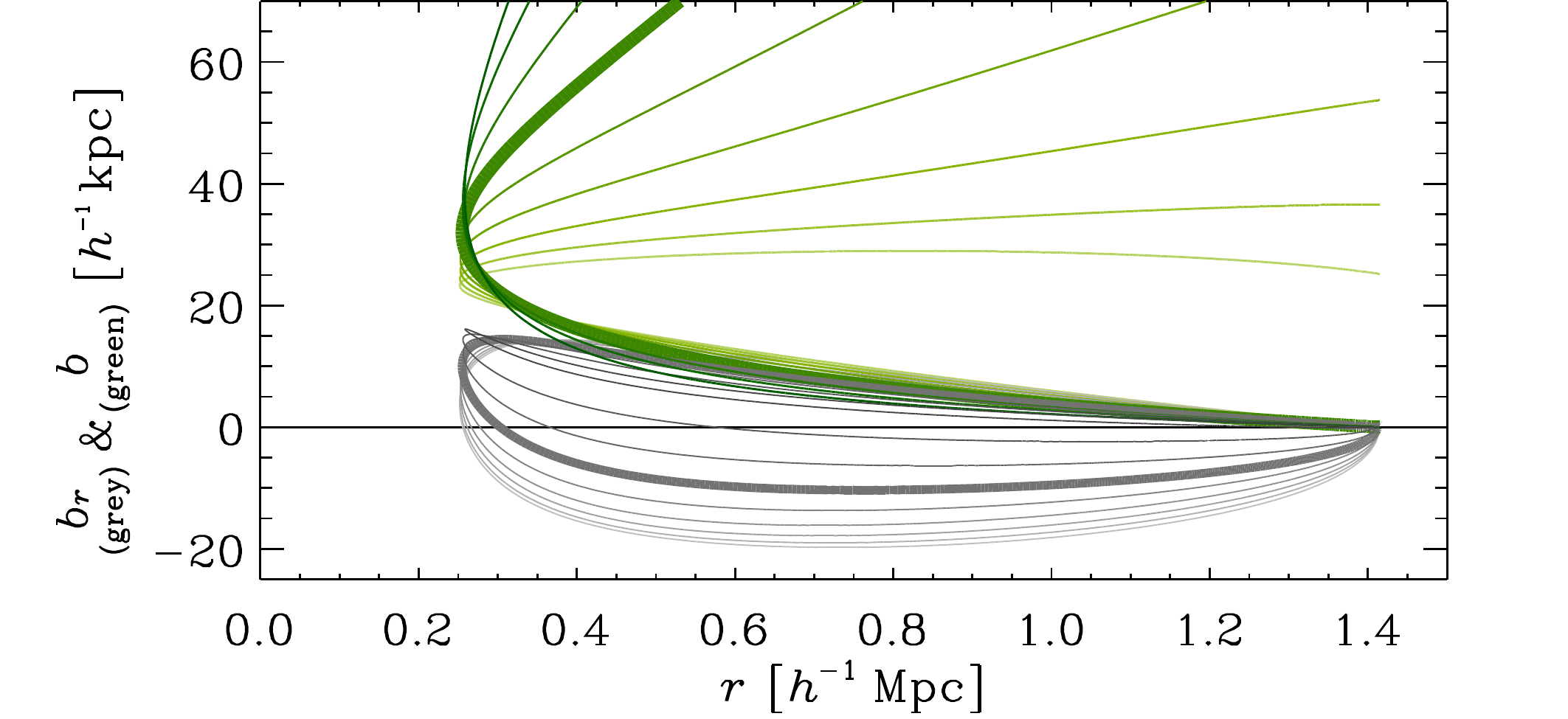}
 \caption{Varying the power law slope of the density profile
 $\rho\propto r^\gamma$ between $\gamma=-1$ (light) and $\gamma=-2.6$ (dark) 
 in steps of 0.2, for a fixed cluster mass and bullet density, while adjusting the 
 bullet's initial velocity to also ensure a constant impact parameter.
 Notice that $\partial b/\partial\gamma$ is small around $\gamma=-2$ (thick line).
 To avoid confusion, the orbits are only traced up to the second
 periapsis (point of closest approach), 
 and bulleticity to the second apoapsis. \label{fig:power}}
\end{figure}

\begin{figure}
 \includegraphics[width = 230pt]{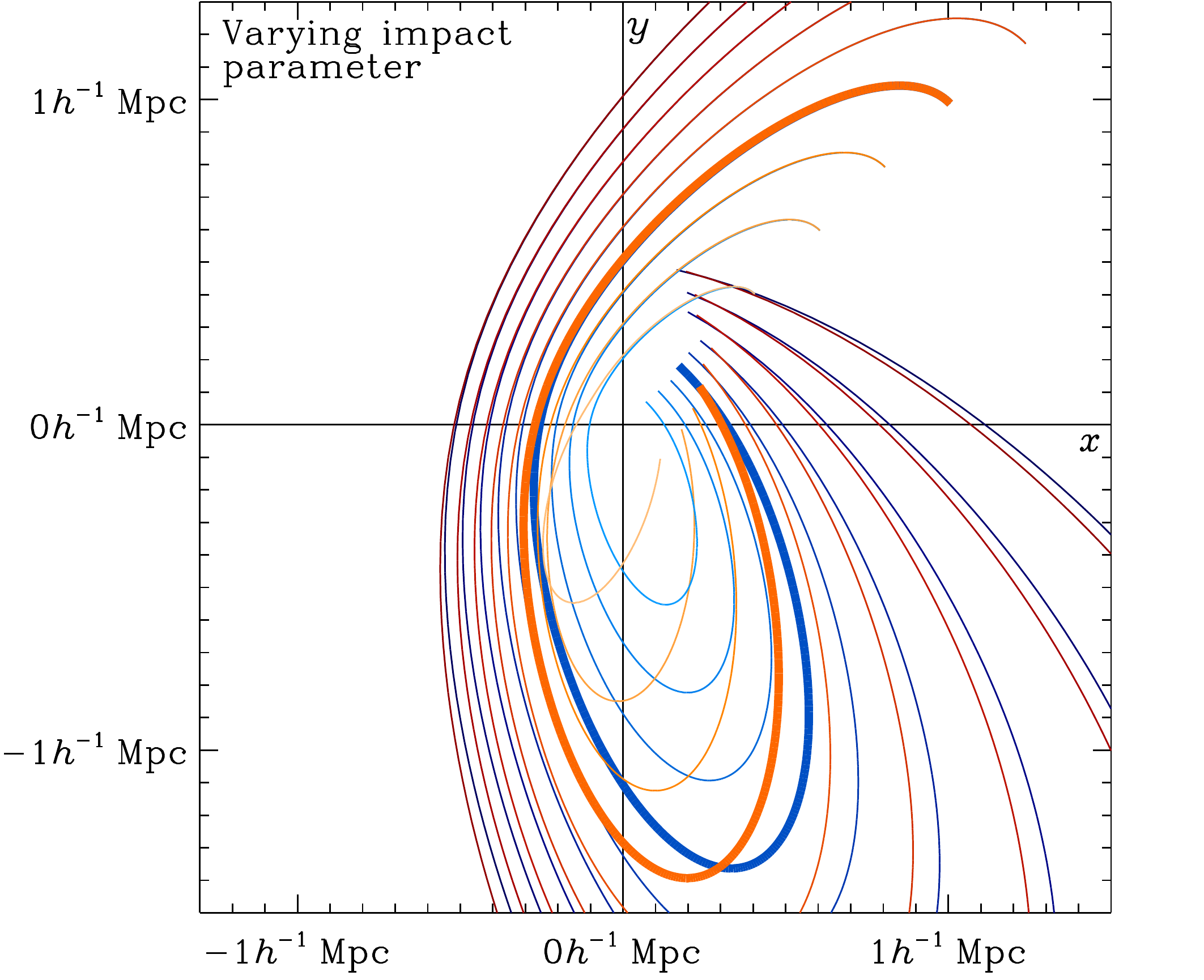}
 \includegraphics[width = 230pt]{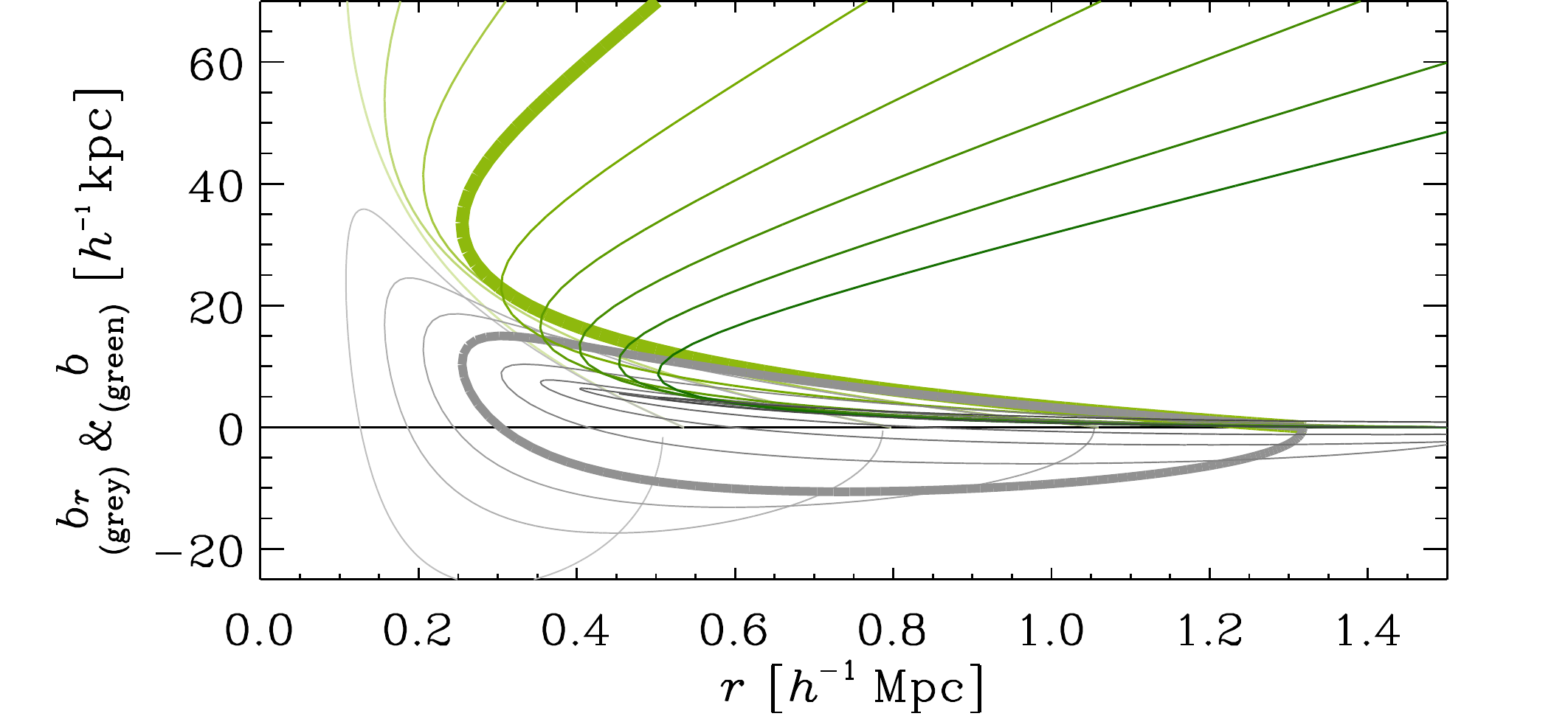}
 \caption{Changing the impact parameter \com{from $100$ to $500\,h^{-1}\,\mathrm{kpc}$
 in steps of $50\,h^{-1}\,\mathrm{kpc}$} (while keeping the impact
 velocity and the cluster profile fixed). \com{The default
 $250\,h^{-1}\,\mathrm{kpc}$ orbit is shown as a thick line.
 The orbit clearly has a more dramatic influence than the cluster's density
 profile (Figure~\ref{fig:power}),} so building a theoretical prediction for real clusters
 will clearly require an integration over initial phase space conditions. \label{fig:impact}}
\end{figure}


The 2D orbit of a bullet with non-zero impact parameter \com{but in the plane of
the sky is illustrated in Figure~\ref{fig:orbit}. In this calculation, we assume
a cluster of mass $10^{14}\,M_\odot$ within $0.8\,h^{-1}\,\mathrm{Mpc}$ and
adopt a value of $\rho_b=10^{6}\,M_\odot/(h^{-1}\,\mathrm{kpc})^3$ for the
baryons; the resulting bulleticity is inversely proportional to this value.} To
show the bulleticity more clearly, the trajectory is extended to include several
passes through the cluster, during each of which the separation of dark and
baryonic matter visibly increases. At each point along the mean trajectory, the
green line in the bottom panel shows the absolute bulleticity $b(r)$. During
the first infall, the separation gradually increases, with baryons lagging
behind dark  matter as in the 1D case. On each subsequent pass, the two
components separate at large radii, as they  follow different trajectories, but
return to each other as they near the cluster core.

\begin{figure}
 \begin{center}
  \includegraphics[scale=0.5]{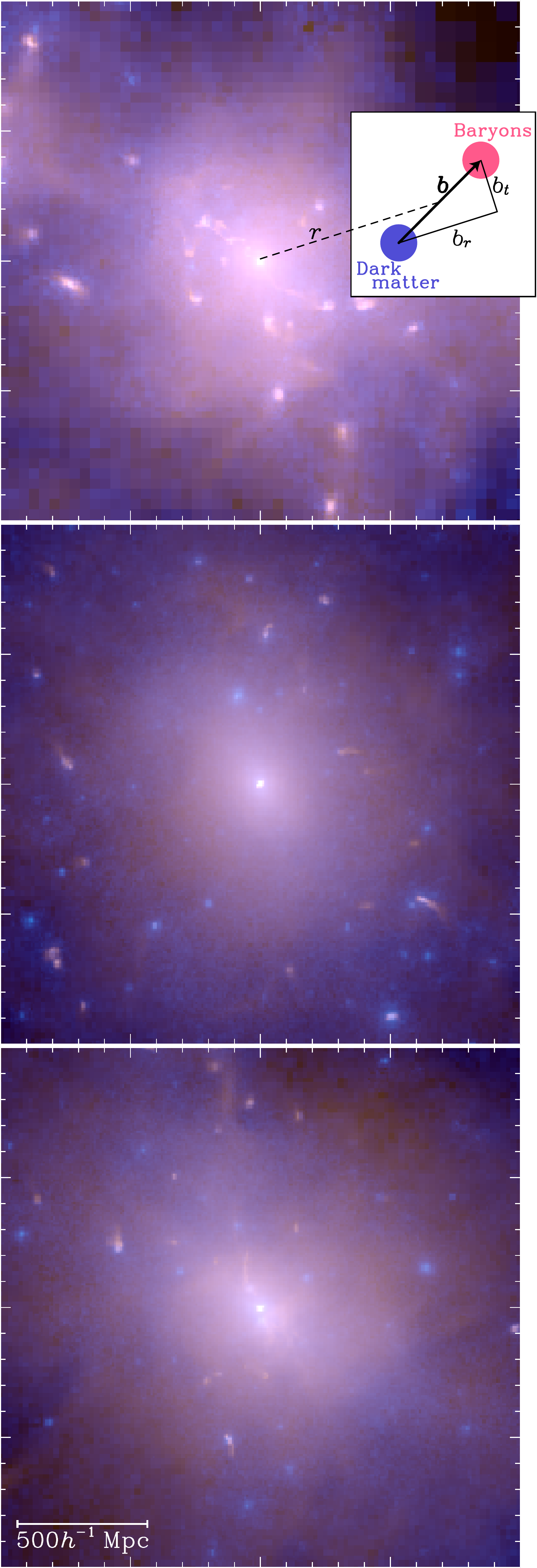}
 \end{center}
 \caption{Full hydrodynamical simulations of massive clusters at redshift $z=0.6$.
 Blue shows total projected mass (dominated by dark matter) and red shows X-ray emission from baryonic gas.
 The preferential trailing of gas due to pressure from the ICM, 
 and its consequent separation from the non-interacting dark matter, is apparent in much of the
 infalling substructure. \label{fig:hydro_images}}
\end{figure}

The bulleticity vector can be conveniently written in terms of components 
that are radial $b_r$ and azimuthal $b_t$ with respect to the centre of the 
cluster, such that
\begin{equation}
\bb=b_r\er+b_t\et ~,
\end{equation}
where $\er$ and $\et$ are unit vectors. \com{This is illustrated
in figure \ref{fig:hydro_images}.
While the radial and tangential bulleticity components are smaller than the
absolute bulleticity $b=|\bb|$, the crucial point is that they are signed.
When averaging over all possible viewing angles or the phase space of possible
bullet orbits}, the mean tangential separation averages to zero (it can be
easily demonstrated in this case by switching the handedness of the orbit to
produce a mirror image of Figure~\ref{fig:orbit}). \com{Such} symmetries ensure
that, for a large sample of bullets, $\langle b_t\rangle\equiv0$. The remaining,
non-zero component of radial  bulleticity $b_r$ is shown in
Figure~\ref{fig:orbit} as a grey line.

\com{We thus propose two complementary bulleticity estimators. The absolute
bulleticity $b$ is positive definite, so is likely to provide the first
detection of a difference in the behaviour of dark and baryonic matter at
relatively large S/N. The radial bulleticity $b_r$ has a smaller signal, but
subsequent measurements from a large cluster sample will benefit from its
corresponding statistical check for systematics, $b_t$. This latter combination
should eventually provide the method's cleanest constraints on the interaction
cross-section of dark matter.}

\subsection{Robustness to astrophysical variation}

The distribution of mass in the cluster affects the orbits of a bullet.
As shown in Figure~\ref{fig:power}, dark matter precesses around a cuspy
mass (converging to Keplerian orbits in the limit of a point-like central mass,
with the mass at one focus of the ellipse), but has  constant elliptical orbits
around a cored cluster, with the centre of the mass at the centre of the
ellipse. Importantly however, for a fixed impact parameter, the bulleticity of
substructure on its first pass through a cluster  changes by less than 5\% for a
wide range of cluster profiles. This lack of influence means that, to first
order, uncertainty about the unknown mass distribution can be ignored, and one
can freely average results from a large ensemble of clusters. It also means that
bulleticity measurements are unlikely to constrain the density profile of galaxy
clusters. However, other methods are expected to constrain this independently
\citep{massey10}.


Conversely, varying a bullet's impact parameter \com{does affect its subsequent
bulleticity. As shown in Figure~\ref{fig:impact}, baryons with a low impact
parameter pass through more of the intra-cluster medium, experience greater
buoyancy, and exhibit larger bulleticity after periapsis. Bullets with a low
impact velocity accumulate a greater impulse from buoyancy, so also have a
higher bulleticity -- especially in the tangential direction, i.e.\  $b_t\gg
b_r$. Deriving quantitative predictions about the level of bulleticity expected
in real clusters will therefore require integrations over the phase space of
initial conditions and subsequent orbits in a typical cosmology, as well as
detailed modelling of the growth and early infall of substructure. Such analysis
is beyond our simple  physical model. To do this properly, we shall now switch
to full hydrodynamical calculations, which also automatically include the more
subtle physical effects that we have disregarded so far.}

\section{Full hydrodynamical calculation} \label{sec:hydro}

\subsection{Properties of the simulations}

We shall now measure the bulleticity signal in realistic, high-resolution
hydrodynamical  simulations of massive clusters embedded in a standard
cosmological model. The model is flat, with parameters $\Omega_{\rm
m}=1-\Omega_{\Lambda}=0.3$, $\Omega_{\rm b}=0.04286$, $h=0.7$ and
$\sigma_8=0.9$, where the Hubble constant is defined as
$100\,h{\rm\,km\,s^{-1}\,Mpc^{-1}}$ and  $\sigma_8$ is the mass variance within
spheres of radius $8\,h^{-1}\,$Mpc. All distances are expressed in comoving
coordinates.

We performed our simulations  with the Adaptive Refinement Tree (ART)
$N$-body$+$gasdynamics code \citep{kra99,kra02}, which uses adaptive refinement
in space and time, and (non-adaptive) refinement in mass \citep{klypin01} to
resolve high dynamic ranges. The spatial resolution in the cores of halos is
$\sim 6\,h^{-1}\,$kpc and the particle mass is 3--9$\times
10^{8}\,h^{-1}\,M_{\odot}$. The  formation of galaxy clusters is followed from
cosmological initial conditions through later properties  of the intracluster
medium including gas cooling and star formation (CSF). Astrophysical processes
in our  simulations include metal enrichment and thermal feedback due to Type Ia
and Type II supernovae, self-consistent advection of metals,
metallicity-dependent radiative cooling and UV heating due to a  cosmological
ionising background. Potentially relevant physical processes excluded from the
simulations  are active galactic nuclei bubbles, magnetic fields, and cosmic
rays, although these are most important  in the innermost cluster regions, which
we shall exclude anyway. More details about our simulations are available in
\citet[][b]{nagai07a}.

We realise simulated 2D observations of $16$ independent clusters at two
redshifts $z=0.6$ and $z=0$ by  projecting snapshots of the dark matter density,
gas density and gas temperature along three orthogonal axes. This produces $48$
cluster realisations at each redshift, some of which are illustrated in
Figure~\ref{fig:hydro_images}. \com{Like the simulations by \cite{powell09},
ours show complex interactions of substructure. Mock {\em Chandra} X-ray imaging
was used to visually classify each projection as either} relaxed or unrelaxed
\citep{nagai07b}. The cluster masses $M_{500}$ span a range of about an order of
magnitude centered on $10^{14}\,M_\odot$, and Table~\ref{tab:properties} lists
the mean mass $M_{500}$, size $r_{500}$ and number of bullets
$N_\mathrm{bullet}$ in several sub-samples, where the extent of the cluster
defined to be the sphere in which the mean density is 500 times the critical
density of the Universe at that epoch.

\begin{table}
 \centering
 \caption{Mean properties of the simulated clusters.}
 \begin{tabular}{lcccc}
  \hline
   ~ & $N_\mathrm{cluster}$ & $N_\mathrm{bullet}$ & Mass $M_{500}$            &
   Size $r_{500}$ \\
   ~ & ~                    & ~                   & [$10^{14}\,M_\odot$] & [$h^{-1}\,$Mpc] \\
  \hline
  \multicolumn{5}{l}{Redshift $z=0.6$}\\
  All         & 48 & 1142 & $1.2\pm 0.2$ & $0.75\pm 0.05$ \\
  Relaxed     & 16 &  303 & $1.6\pm 0.5$ & $0.79\pm 0.09$ \\
  Unrelaxed   & 32 &  839 & $1.0\pm 0.2$ & $0.68\pm 0.04$ \\
  \hline
  \multicolumn{5}{l}{Redshift $z=0$}\\
  All         & 48 & 1079 & $2.9\pm 0.6$ & $0.73\pm 0.06$ \\
  Relaxed     & 27 &  453 & $2.0\pm 0.6$ & $0.66\pm 0.07$ \\
  Unrelaxed   & 21 &  626 & $3.6\pm 0.9$ & $0.79\pm 0.08$ \\
  \hline\label{tab:properties}
  \vspace{-6mm}
 \end{tabular}
\end{table}


%

\subsection{Analysis of the simulations}

\begin{figure}
 \includegraphics[width = 230pt]{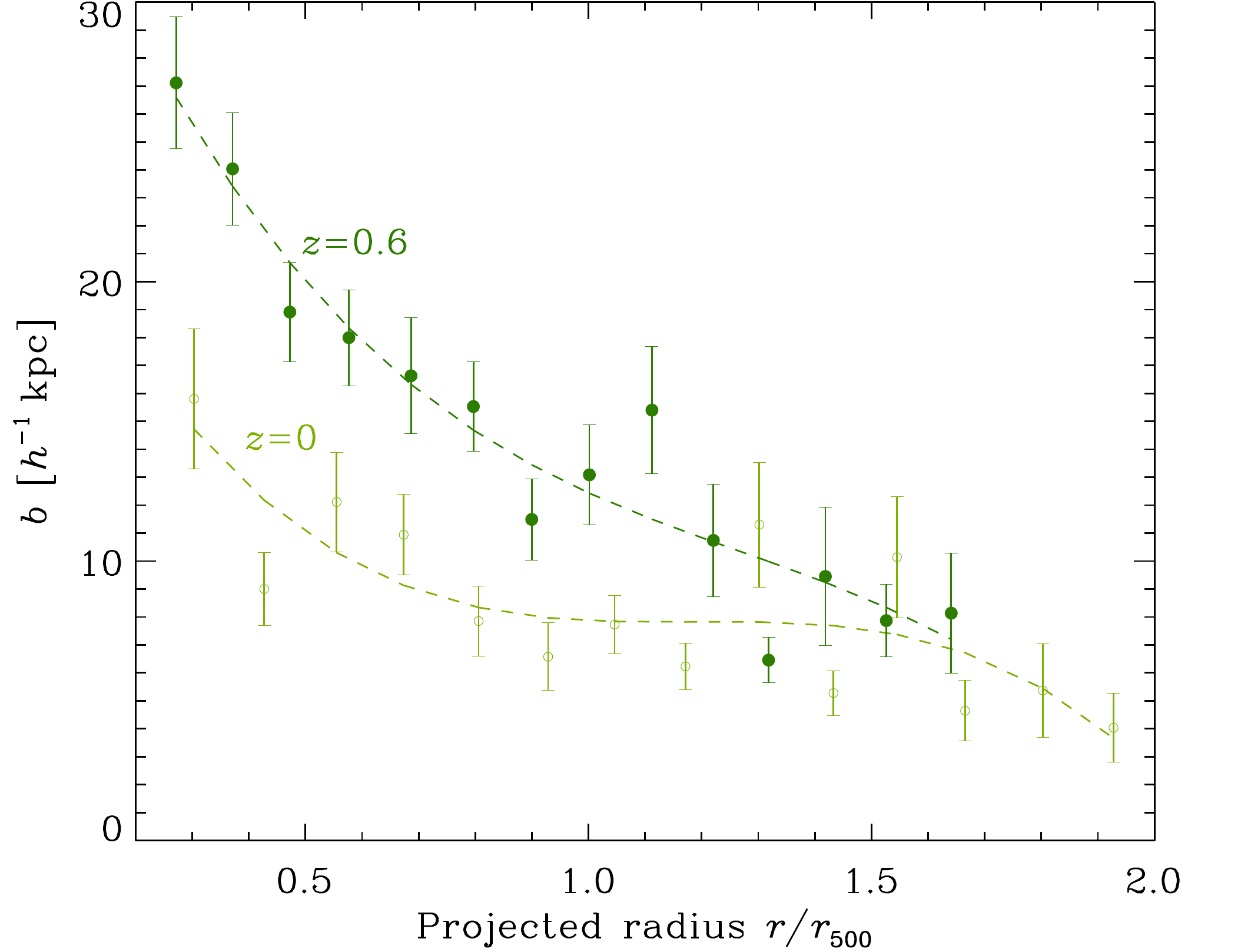}
 \caption{Measurements of absolute bulleticity from full hydrodynamical simulations. 
 Error bars show $1\sigma$ errors.
 Dashed lines are best-fitting cubic polynomials to guide the eye. 
 As expected for orbits in which only the first crossing has been completed, bulleticity
 increases as the infalling group falls towards the cluster centre and the gas is preferentially
 retarded.\label{fig:hydro_b}}
\end{figure}

\begin{figure}
 \includegraphics[width = 230pt]{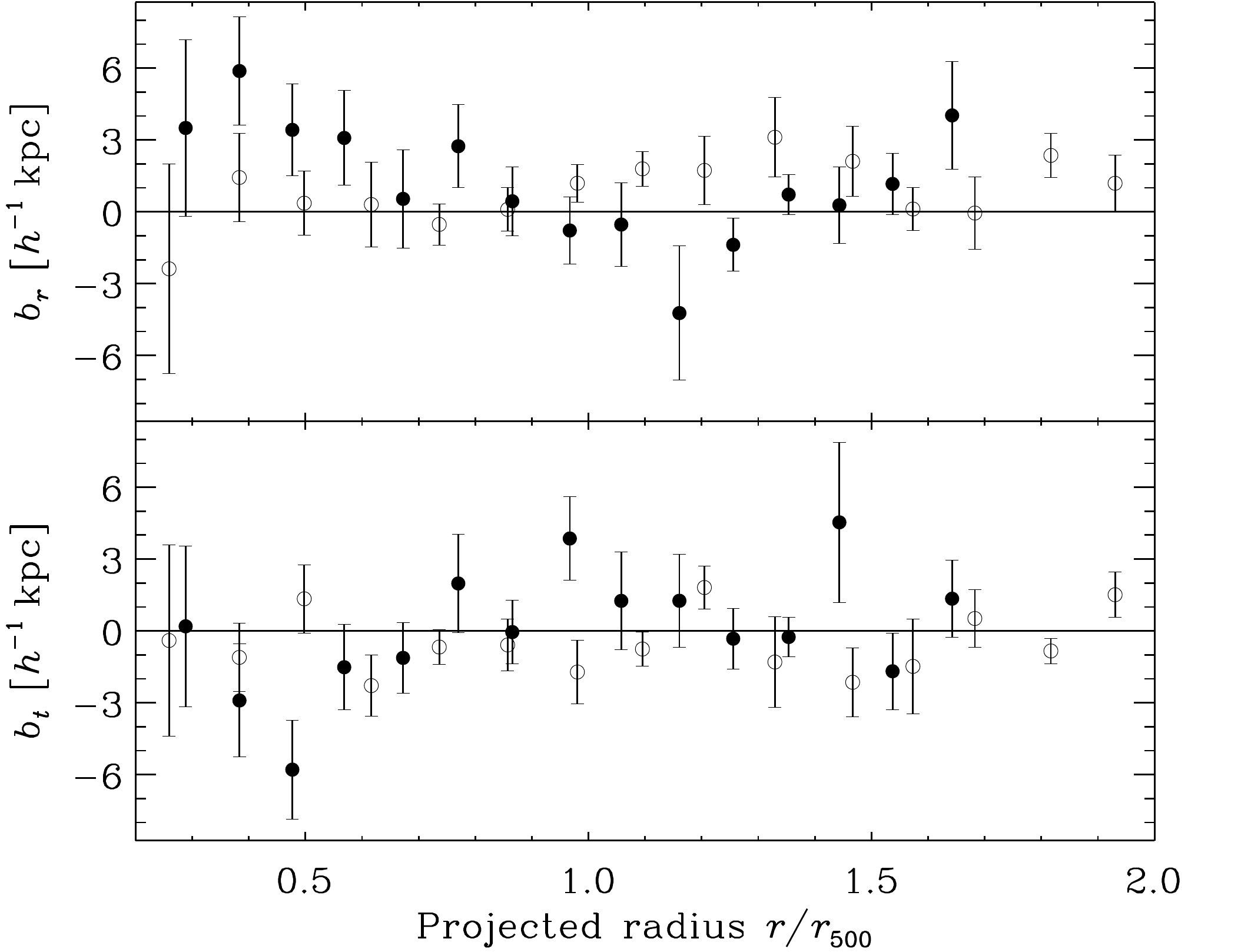}
 \caption{Measurements of \com{radial and tangential} bulleticity from full hydrodynamical simulations. 
 \com{These quantities are signed so, once averaged over many clusters, the 
 values are smaller than those in Figure~\ref{fig:hydro_b}.}
 Filled circles show data from redshift $z=0.6$ and open circles show data from $z=0$. 
 The only significant detection is of radial bulleticity
 $b_r(r<r_{500})=1.7\pm 0.7\,h^{-1}\,$kpc at $z=0.6$.  
 \com{The mean} tangential bulleticity is consistent with zero, 
 demonstrating its utility as a control test in this harder but potentially more discriminating measurement. \label{fig:hydro_brt}}
\end{figure}

We apply unsharp masking to maps of both the lensing mass and the baryonic gas 
density squared times the square root of the temperature (equivalent to the
X-ray emission only if it is due to thermal bremsstrahlung, but the rest of our
analysis is completely robust to this simplification).  We then find the
best-fit Gaussian to every local maximum via the iteration in size and position
adopted by \citet{rrg}. This algorithm is robust but may not be optimal in the
regime of real observational  noise and beam smearing. Even in our noiseless
simulations, measurements near the cluster core are  hindered by the steep
background gradient from central emission, particularly in the X-ray signal,
which  pulls the best-fitting peak inward. More sophisticated algorithms might
prevent this, such as simultaneous fitting of all the sources
\citep[c.f.][]{marshall06}. However, we circumvent the issue by ignoring the
(very few) peaks within $200\,h^{-1}\,\mathrm{kpc}$ of the cluster centre. We
cross-match X-ray peaks to their nearest lensing peak projected within
$0.5\,h^{-1}\,\mathrm{Mpc}$, then apply the match the other way around and keep 
only uniquely defined pairs. 

\com{Figures~\ref{fig:hydro_b} and \ref{fig:hydro_brt} show the observed mean
separation between the centres of matched X-ray and lensing peaks, projected
within $0.5\,h^{-1}\,$Mpc on the sky and averaged over all subpeaks. The radial
and tangential components $b_r$ and $b_t$ can be positive or negative, so they
partially cancel upon averaging, while $b\equiv\sqrt{b_r^2+b_t^2}$ is positive
definite for all bullets. As in \S\ref{sec:toymodel}, we define the centre of
the bullet as the mean of the X-ray and lensing positions, and we compute the
components of the bulleticity vector with respect to the direction towards the
global minimum of the projected gravitational potential}. As expected from the
simple model in \S\ref{sec:toymodel}, the absolute bulleticity increases towards
the centre of the cluster, reaching $26\,h^{-1}$\,kpc\,$\approx 4\arcsec$ at
$r=0.3\,r_{500}$ and $z=0.6$. The \com{normalisation} is independent of the cluster mass (at
both redshifts), within a $1\sigma$ uncertainty of  $4\,h^{-1}$\,kpc per order
of magnitude in mass. \com{The measurements are well fit by cubic polynomials
\be
b(r)=\frac{\Delta\sigma}{\pi r_b^2}
\left( ~b_0 + b_1 r + b_2 r^2 + b_3 r^3 ~\right)~,
\label{eqn:bpoly}
\ee
where the coefficients $b_i=\{38.2, -51.6, 35.0, -9.2\}$ at $z=0.6$ and
$b_i=\{24.7, -43.0, 36.4, -10.3\}$ at $z=0$, assuming a fiducial model in which
the prefactor is unity. If the polynomial order is increased, the quartic
coefficients are an order of magnitude lower.}



The radial and tangential bulleticity signals are an order of magnitude lower
and noisier. \com{The only statistically significant detection is that} 
$\langle b_r\rangle=1.7\pm 0.7\,h^{-1}\,\mathrm{kpc}$ within $r_{500}$
at $z=0.6$. This too appears independent of cluster mass, and all other
measurements are consistent with zero. In particular, that the tangential
bulleticity signal is always consistent with zero demonstrates its utility as a
\com{control test for any observational systematics.}

\subsection{Dependence upon redshift}

The bulleticity signal decreases between the two cosmic epochs we have studied.
\com{An explanation for this, in agreement with our predictions from the simple
model of \S\ref{sec:toymodel}, is that the masses of clusters are considerably
lower at $z=0.6$. With a correspondingly lower infall velocity, the  baryons
therefore experience a larger impulse from buoyancy.} Furthermore, high redshift
clusters are also generally more disturbed, have more substructures per unit
mass, and have larger offsets between central dark matter and gas peaks.
However, the projected angle of $b$ on the sky {\em increases} for nearby
clusters. Were the absolute bulleticity to decline linearly with proper time
between the values measured at redshifts $z=0.6$ and 0, the expectation of
$b(0.3\,r_{500})$ would rise in angular size to $\sim10\arcsec$ at $z=0.1$ and 
$\sim16\arcsec$ at $z=0.05$. 

\com{Clusters at low redshift will make the best targets for observation. As well
as their increasing bulleticity signal, nearby clusters also provide a more
optimal geometry for gravitational lensing. The common misconception to the
contrary may have arisen from \cite{hamana04}, who detected peaks using a
matched filter of fixed $1\arcmin$ size. Nearby clusters appear larger than this
on the sky, so their fixed filter produces a diminishing signal below
$z\sim0.3$. A better matched filter reveals that the signal is larger, but
merely spread thin \citep{kubo07,kubo09}. Finally, massive, low redshift
clusters should contain more substructure. That our sample includes a similar
amount of substructure at all redshifts is probably a selection effect because
our high redshift clusters are the most massive progenitors of present-day
structures, so are growing rapidly.}

\subsection{Dependence upon environment}

None of the bulleticity signals show a statistically significant dependence upon
the cluster's apparent dynamical state. However, at redshift $z=0.6$~(0),
unrelaxed clusters yielded about 1.4~(1.8) times as many bullets per cluster as
relaxed clusters, despite having slightly lower masses at $z=0.6$. For this
measurement, unrelaxed clusters might therefore provide more profitable
targets. 

We also tried stacking bulleticity measurements from different clusters in terms
of the bullets' absolute distance from the cluster centre, rather than as a
fraction of $r_{500}$. Because the cluster sizes vary little within our sample,
the qualitative result does not change.

In particularly crowded regions of clusters, it would be possible to mismatch
pairs of dark matter and  baryonic projected peaks from different substructures.
However, this confusion effect will dampen the  measured bulleticity and, if it
is a function of cluster radius, it will be most pronounced towards the centre,
where the separation is largest, and the density of bullets is highest. If the
aim is purely to  detect bulleticity $b$ in order to prove the existence of dark
matter, this effect will therefore only  make the measurement more difficult
rather than producing a spurious signal.

\section{Practicality of a real measurement} \label{sec:practicality}

\com{So far, we have maximally exploited our computationally expensive simulations by
not adding noise to our mass or X-ray maps. We shall now consider the likelihood
of and practical issues that will be faced by any real measurement of
bulleticity.

Baryonic substructure is frequently seen in deep {\em Chandra} imaging
\citep[e.g.][]{gastaldello09,randall09}. Dark matter substructure can be mapped
efficiently via strong gravitational lensing or flexion
\citep{goldberg05,bacon06}, which probes gradients in the mass distribution and
is therefore more sensitive to small mass peaks along a line of sight than weak
shear. \cite{coe10} resolved ten previously unknown subpeaks at various radii
within (unrelaxed) cluster A1689 using strong lensing, and
\cite{leonard07,leonard10} resolved four using flexion. Both of these
measurements were made independently of the distribution of light.

Combined X-ray and gravitational lensing observations have already revealed a
separation of baryons from dark matter in two real systems undergoing major
mergers. In the bullet cluster 1E~0657-56, \cite{clowe06} measured separations
of $b=49.3\arcsec=152\,h^{-1}\,\mathrm{kpc}$ (main cluster) and
$46.1\arcsec=142\,h^{-1}\,\mathrm{kpc}$ (bullet). In MACS~J0025.4-1222,
\cite{bradac08} measured separations of dark matter from the central gas peak of
$b=49.3\arcsec=228\,h^{-1}\,\mathrm{kpc}$ (SE clump) and
$30.1\arcsec=139\,h^{-1}\,\mathrm{kpc}$ (NW clump). While these measurements
have small errors, it is their interpretation that remains difficult. The
extreme disruption of these systems has removed any well-defined global
potential minimum, so it is difficult to place the substructure at a
well-defined radius $r$ or to split the bulleticity into components $b_r$ and
$b_t$. Minor merger events will be more usual targets for bulleticity.}


Noise in a real measurement -- whether due to a finite exposure time for X-ray
observations or a finite source density of lensed background galaxies -- will
create scatter in the measured peak positions. This will emerge as a constant
minimum  $b$ signal. Indeed, this effect is tentatively seen as a value around
$6\,h^{-1}\,\mathrm{kpc}$ at large $r$ in Figure~\ref{fig:hydro_b}, which
coincides with the resolution limit of the simulations. \com{A resolution limit
will be especially problematic in clusters at high redshift, although
statistical techniques more sophisticated than an offset between peaks will
inevitably help. For example, the cross-correlation between the full lensing and
X-ray maps could be measured. Most importantly, measurements of the (signed)
radial and tangential bulleticity from a large cluster sample will beat down
noise on $\langle  b_r\rangle$, and ensure that $\langle b_t\rangle\rightarrow
0$.

To predict the observable bulleticity signal, we assume that substructure
positions can be resolved to $6\,h^{-1}\,\mathrm{kpc}$, which is achievable with
{\em Chandra} at $z<0.3$. At radii where the expected bulleticity
signal~(\ref{eqn:bpoly}) is resolved, we assume that the observed r.m.s.\ error
$\sigma_b(r)$ falls from the baseline of Figure~\ref{fig:hydro_b} as
$1/\sqrt{N_b}$, where $N_b$ is the number of bullets. Once the approved
Multi-Cycle Treasury programme `Through a lens, darkly' (P.I. Postman) has been
performed, the HST/ACS imaging archive will include $\sim50$ clusters between
redshifts $z=0.1$ and $0.2$. If these clusters contain just $\sim100$
substructure peaks, and even excluding measurements of those within the central
$200\,h^{-1}\,\mathrm{kpc}$, they should provide a detection of $b(r)$ at a
signal to noise ratio of $\sim11$ 
when integrated over scales $6\,h^{-1}\,\mathrm{kpc}<r<2r_{500}$. The Euclid survey \citep{massey04,euclid} should yield
a similar detection significance for $b_r(r)$ and $b_t(r)$, even if only one
substructure peak is identified in each of its $\sim40,000$ clusters within the
same redshift range.


We can estimate the tightness with which such measurements will constrain the
dark matter interaction cross-section using the Fisher information matrix
\citep{tegmark97}. We again adopt the best fit models of $b(r)$ from
equation~(\ref{eqn:bpoly}), interpolating linearly with proper time between
$z=0.6$ and $z=0$, and baseline observational noise $\sigma_b(r)$ around that
shown in Figure~\ref{fig:hydro_b}. We assume that the bulleticity in
radial bins is uncorrelated, and note that the dependence on $\Delta\sigma$ is
in the mean of  $b(r)$ (not the covariance). In this case, the one-parameter
Fisher information is
\ba
F&=&\sum_{\rm bullets}\sum_{r} \frac{1}{\sigma^2_b(r)}
    \left(\frac{\partial b(r)}{\partial \Delta\sigma}\right)^2\nonumber\\
 &=&\sum_{\rm bullets} \frac{1}{(\pi r_b^2)^2}
    \sum_{r} \frac{\left( ~b_0 + b_1 r + b_2 r^2 + b_3 r^3~\right)^2}{\sigma^2_b(r)~~~}~.
\label{eqn:fisher}
\ea
We have summed equally over $r$ bins where $r>200\,h^{-1}\,\mathrm{kpc}$ and 
$b(r)>6\,h^{-1}\,\mathrm{kpc}$, but this could be generalised to incorporate a
more sophisticated weight function that raises the overall signal to noise. For
the observational scenarios described above, $F(\pi r_b^2)^2\sim100$ for the 
HST archive, and $\sim30,000$ for Euclid. 

Notably, equation~(\ref{eqn:fisher}) does not depend on the fiducial value of 
$\Delta\sigma$. With a more comprehensive suite of hydrodynamical simulations to
more accurately model the behaviour of baryonic substructure, we could therefore
directly interpret constraints on $\Delta\sigma$ as those on $\sigma$. Assuming
typical bullets of mass $m_b\sim5\times10^{12}$ and radius
$r_b\sim10\,h^{-1}\,\mathrm{kpc}$, we thus predict 
68\% confidence limits 
of
\be
\Bigg(\frac{\sigma}{m}\Bigg)=\frac{1}{m_b\sqrt{F}}\sim
\begin{cases} 
1\times10^{-25}\,\mathrm{cm}^2/\mathrm{GeV} & \text{for HST} \\
6\times10^{-27}\,\mathrm{cm}^2/\mathrm{GeV} & \text{for Euclid.}
\end{cases}
\ee
Compare this to 68\% confidence limits from the bullet cluster of
$\sigma/m<1.25\,\mathrm{cm}^2/\mathrm{g}=2\times10^{-24}\mathrm{cm}^2/\mathrm{GeV}$
or $\sigma/m<0.7\,\mathrm{cm}^2/\mathrm{g}$ assuming that the main cluster and sub-cluster had similar mass-to-light ratios prior to the merger
\citep{randall08}.

Even with these fairly conservative estimates, we expect a bulleticity analysis
of the HST archive to produce constraints on $\sigma$ similar to or tighter than
the bullet cluster. Furthermore, such constraints are potentially unlimited by
the uncertainty in orbital parameters for any single object. The fundamental
strength of our statistical method is the trivial way in which it can then be
extended to exploit larger surveys like Euclid. Bulleticity thus offers a path
towards ever more discriminating measurements, even if  individual extreme
merger events turn out to be rare \citep{shan10}.}


\section{DISCUSSION} \label{sec:discuss}

We have defined a measure of dark matter-baryonic matter separation
`bulleticity' that takes contributions from every detected substructure peak in
every massive cluster. Any non-zero bulleticity measures the difference between
the interaction cross-sections of dark and baryonic matter. This interesting new
test can be understood via the intuition of a simple model, and we have also
measured \com{the expected value of the bulleticity signal} using full
hydrodynamical simulations of $\Lambda$CDM clusters. 

\com{A conservative estimate of currently available data and analysis techniques
suggests that there should be enough information in clusters from the HST
archive to detect bulleticity at a signal to noise greater than $10$. With
further hydrodynamical simulations to interpret the absolute level of the
signal, this could yield independent constraints on the interaction
cross-section of dark matter at a similar or tighter level than the bullet
cluster. The real strength of this method is the way in which it can
subsequently exploit large future surveys, free of biases from individual
systems and in a trivially expandable way. An ambitious, all-sky survey could
thus rival constraints from particle physics experiments \citep{feng10}. For a
targetted survey, the ideal targets would include massive, low redshift, and
possibly unrelaxed clusters. Most crucially, bulleticity measurements will be
obtained in a physical regime unapproachable in terrestrial laboratories, and
may ultimately provide the best test for any velocity dependence of the dark
matter interaction cross-section.}\\

\com{As a final tantalising prospect, we note that} the positions of galaxies
provide a third (and more easily measured) observable. \com{Indeed,
\cite{randall08} derived tighter constraints on the dark matter cross section by
comparing the post-collision locations of dark matter and galaxies (rather than
dark matter and gas). To first order, galaxies pass straight through each other
unimpeded because of the separation between them, and between the stars in each
galaxy. The vector from dark matter to galaxies is therefore a `bulleticity'
measured around a fiducial model of $\sigma=0$, so it provides a more direct
measurement of nonzero interaction cross-section. There are still potential
complications to this picture. For example, \cite{russell10} found a group of
galaxies in A2146 leading the X-ray emission as expected, but the brightest
cluster galaxy lags behind it. Complex baryonic physics can also affect
observations:} \cite{cortese07} showed that stars lead gas in galaxies that are
merging into clusters, but that new star formation can also be triggered in the
gas, with the new stars only gradually falling forward into the main galaxy.
\com{If such effects can be theoretically modelled, the most practical tool for
this measurement is likely to be strong gravitational lensing. If the positions
of peaks in a standard {\it LensFit} mass reconstruction
\citep[e.g.][]{richard10} were allowed to float instead of being tied to the
positions of galaxies, their measured offsets would be precisely this new
bulleticity.}

\section*{Acknowledgments}

The authors would like to thank Patrick Simon, Ken Rice, Phil Marshall, Cathie
Clarke, Justin Read, Johan Richard and James Taylor for conversations that
spurred the development of this paper. It also benefitted greatly from helpful
suggestions by the anonymous referee. RM is supported by STFC Advanced
Fellowship \#PP/E006450/1 and ERC grant MIRG-CT-208994. TK is supported by STFC
rolling grant RA0888. DN acknowledges the support of Yale University and NSF 
grant AST-1009811. 


\bsp

\label{lastpage}

\end{document}